\documentclass[twocolumn,trackchanges]{aastex701}

\usepackage{tabularx}
\usepackage{longtable}
\usepackage{extarrows}
\usepackage{supertabular}
\usepackage{threeparttablex}
\usepackage{graphicx}% Include figure files
\usepackage{dcolumn}% Align table columns on decimal point
\usepackage{bm}% bold math
\usepackage{amsmath}
\usepackage{overpic}
\usepackage{booktabs}%
\usepackage{makecell}
\usepackage{enumerate}
\usepackage{multirow}
\usepackage{rotating}[figuresright]
\makeatletter

\newcommand{\Rmnum}[1]{\expandafter\@slowromancap\romannumeral #1@}
\makeatother
\usepackage{hyperref}
\usepackage{xcolor}
\begin{document}

\title{\texttt{BREAKFAST}: A Framework for general joint BA duty and follow-up guidance of multiple $\gamma$-ray monitors}

\correspondingauthor{Shao-Lin Xiong, Yue Huang}
\email{xiongsl@ihep.ac.cn, huangyue@ihep.ac.cn}

\author[0009-0008-8053-2985]{Chen-Wei Wang}
\affil{State Key Laboratory of Particle Astrophysics, Institute of High Energy Physics, Chinese Academy of Sciences, Beijing 100049, China}
\affil{University of Chinese Academy of Sciences, Chinese Academy of Sciences, Beijing 100049, China}
\email{cwwang@ihep.ac.cn}

\author[0000-0002-8097-3616]{Peng Zhang}
\affil{State Key Laboratory of Particle Astrophysics, Institute of High Energy Physics, Chinese Academy of Sciences, Beijing 100049, China}
\affil{College of Electronic and Information Engineering, Tongji University, Shanghai 201804, China}
\email{zhangp97@ihep.ac.cn}

\author[0000-0002-4771-7653]{Shao-Lin Xiong*}
\affil{State Key Laboratory of Particle Astrophysics, Institute of High Energy Physics, Chinese Academy of Sciences, Beijing 100049, China}
\email{xiongsl@ihep.ac.cn}

\author{Yue Huang*}
\affil{State Key Laboratory of Particle Astrophysics, Institute of High Energy Physics, Chinese Academy of Sciences, Beijing 100049, China}
\email{huangyue@ihep.ac.cn}

\author[0009-0006-5506-5970]{Wen-Jun Tan}
\affil{State Key Laboratory of Particle Astrophysics, Institute of High Energy Physics, Chinese Academy of Sciences, Beijing 100049, China}
\affil{University of Chinese Academy of Sciences, Chinese Academy of Sciences, Beijing 100049, China}
\email{tanwj@ihep.ac.cn}

\author[0009-0002-6411-8422]{Zheng-Hang Yu}
\affil{State Key Laboratory of Particle Astrophysics, Institute of High Energy Physics, Chinese Academy of Sciences, Beijing 100049, China}
\affil{University of Chinese Academy of Sciences, Chinese Academy of Sciences, Beijing 100049, China}
\email{zhyu@ihep.ac.cn}

\author[0009-0008-5068-3504]{Yue Wang}
\affil{State Key Laboratory of Particle Astrophysics, Institute of High Energy Physics, Chinese Academy of Sciences, Beijing 100049, China}
\affil{University of Chinese Academy of Sciences, Chinese Academy of Sciences, Beijing 100049, China}
\email{yuewang@ihep.ac.cn}

\author[0000-0001-8664-5085]{Wang-Chen Xue}
\affil{State Key Laboratory of Particle Astrophysics, Institute of High Energy Physics, Chinese Academy of Sciences, Beijing 100049, China}
\affil{University of Chinese Academy of Sciences, Chinese Academy of Sciences, Beijing 100049, China}
\email{xuewangchen@ihep.ac.cn}

\author[0009-0001-7226-2355]{Chao Zheng}
\affil{State Key Laboratory of Particle Astrophysics, Institute of High Energy Physics, Chinese Academy of Sciences, Beijing 100049, China}
\affil{University of Chinese Academy of Sciences, Chinese Academy of Sciences, Beijing 100049, China}
\email{zhengchao97@ihep.ac.cn}

\author{Hao-Xuan Guo}
\affil{State Key Laboratory of Particle Astrophysics, Institute of High Energy Physics, Chinese Academy of Sciences, Beijing 100049, China}
\affil{Department of Nuclear Science and Technology, School of Energy and Power Engineering, Xi'an Jiaotong University, Xi'an, China}
\email{guohx@ihep.ac.cn}

\author[0000-0002-6540-2372]{Ce Cai}
\affil{College of Physics and Hebei Key Laboratory of Photophysics Research and Application, Hebei Normal University, Shijiazhuang, Hebei 050024, China}
\email{caice@hebtu.edu.cn}

\author[0000-0003-3882-8316]{Yong-Wei Dong}
\affil{State Key Laboratory of Particle Astrophysics, Institute of High Energy Physics, Chinese Academy of Sciences, Beijing 100049, China}
\email{dongyw@ihep.ac.cn}

\author{Jiang He}
\affil{State Key Laboratory of Particle Astrophysics, Institute of High Energy Physics, Chinese Academy of Sciences, Beijing 100049, China}
\email{hejiang@ihep.ac.cn}

\author[0000-0001-5798-4491]{Cheng-Kui Li}
\affil{State Key Laboratory of Particle Astrophysics, Institute of High Energy Physics, Chinese Academy of Sciences, Beijing 100049, China}
\email{lick@ihep.ac.cn}

\author[0000-0003-4585-589X]{Xiao-Bo Li}
\affil{State Key Laboratory of Particle Astrophysics, Institute of High Energy Physics, Chinese Academy of Sciences, Beijing 100049, China}
\email{lixb@ihep.ac.cn}

\author[0009-0004-1887-4686]{Jia-Cong Liu}
\affil{State Key Laboratory of Particle Astrophysics, Institute of High Energy Physics, Chinese Academy of Sciences, Beijing 100049, China}
\affil{University of Chinese Academy of Sciences, Chinese Academy of Sciences, Beijing 100049, China}
\email{liujiacong@ihep.ac.cn}

\author{Xing-Hao Luo}
\affil{Institute for Frontier in Astronomy and Astrophysics, Beijing Normal University, Beijing 102206, China, Department of Astronomy, Beijing Normal University, Beijing 100875, China}
\affil{Department of Astronomy, Beijing Normal University, Beijing 100875, China}
\email{202211998368@mail.bnu.edu.cn}

\author{Xiang Ma}
\affil{State Key Laboratory of Particle Astrophysics, Institute of High Energy Physics, Chinese Academy of Sciences, Beijing 100049, China}
\email{max@ihep.ac.cn}

\author{Rahim Moradi}
\affil{State Key Laboratory of Particle Astrophysics, Institute of High Energy Physics, Chinese Academy of Sciences, Beijing 100049, China}
\email{rmoradi@ihep.ac.cn}

\author{Yang-Zhao Ren}
\affil{School of Physical Science and Technology, Southwest Jiaotong University, Chengdu Sichuan, 611756, China}
\affil{State Key Laboratory of Particle Astrophysics, Institute of High Energy Physics, Chinese Academy of Sciences, Beijing 100049, China}
\email{renyz@ihep.ac.cn}  

\author[0000-0003-0274-3396]{Li-Ming Song}
\affil{State Key Laboratory of Particle Astrophysics, Institute of High Energy Physics, Chinese Academy of Sciences, Beijing 100049, China}
\affil{University of Chinese Academy of Sciences, Chinese Academy of Sciences, Beijing 100049, China}
\email{songlm@ihep.ac.cn}

\author{Ping Wang}
\affil{State Key Laboratory of Particle Astrophysics, Institute of High Energy Physics, Chinese Academy of Sciences, Beijing 100049, China}
\email{pwang@ihep.ac.cn}

\author{Jin Wang}
\affil{State Key Laboratory of Particle Astrophysics, Institute of High Energy Physics, Chinese Academy of Sciences, Beijing 100049, China}
\email{jinwang@ihep.ac.cn}

\author{Bo-Bing Wu}
\affil{State Key Laboratory of Particle Astrophysics, Institute of High Energy Physics, Chinese Academy of Sciences, Beijing 100049, China}
\email{wubb@ihep.ac.cn}

\author[0000-0003-2957-2806]{Shuo Xiao}
\affil{Guizhou Provincial Key Laboratory of Radio Astronomy and Data Processing, 
 Guizhou Normal University, Guiyang 550001, China}
\affil{School of Physics and Electronic Science, Guizhou Normal University, Guiyang 550001, China}
\email{xiaoshuo@gznu.edu.cn}

\author[0000-0001-9217-7070]{Sheng-Lun Xie}
\affil{State Key Laboratory of Particle Astrophysics, Institute of High Energy Physics, Chinese Academy of Sciences, Beijing 100049, China}
\affil{Institute of Astrophysics, Central China Normal University, Wuhan 430079, China}
\email{xiesl@mails.ccnu.edu.cn}

\author[0000-0001-7599-0174]{Shu-Xu Yi}
\affil{State Key Laboratory of Particle Astrophysics, Institute of High Energy Physics, Chinese Academy of Sciences, Beijing 100049, China}
\email{sxyi@ihep.ac.cn}

\author{Xue-Yuan Zao}
\affil{State Key Laboratory of Particle Astrophysics, Institute of High Energy Physics, Chinese Academy of Sciences, Beijing 100049, China}
\affil{University of Chinese Academy of Sciences, Chinese Academy of Sciences, Beijing 100049, China}
\email{zaoxy@ihep.ac.cn}

\author{Xiao-Yun Zhao}
\affil{State Key Laboratory of Particle Astrophysics, Institute of High Energy Physics, Chinese Academy of Sciences, Beijing 100049, China}
\email{xyzhao@ihep.ac.cn}

\author{Li Zhang}
\affil{State Key Laboratory of Particle Astrophysics, Institute of High Energy Physics, Chinese Academy of Sciences, Beijing 100049, China}
\email{zhangli@ihep.ac.cn}

\author[0000-0001-5586-1017]{Shuang-Nan Zhang}
\affil{State Key Laboratory of Particle Astrophysics, Institute of High Energy Physics, Chinese Academy of Sciences, Beijing 100049, China}
\affil{University of Chinese Academy of Sciences, Chinese Academy of Sciences, Beijing 100049, China}
\email{zhangsn@ihep.ac.cn}

\author[0000-0001-5348-7033]{Yan-Qiu Zhang}
\affil{Guizhou Provincial Key Laboratory of Radio Astronomy and Data Processing, 
 Guizhou Normal University, Guiyang 550001, China}
\affil{School of Physics and Electronic Science, Guizhou Normal University, Guiyang 550001, China}
\email{yqzhang@ihep.ac.cn}

\author[0000-0003-2256-6286]{Shi-Jie Zheng}
\affil{State Key Laboratory of Particle Astrophysics, Institute of High Energy Physics, Chinese Academy of Sciences, Beijing 100049, China}
\email{zhengsj@ihep.ac.cn}

\begin{abstract}
With the growing number of $\gamma$-ray monitors in operation, several research teams have adopted a strategy of joint operation and scientific duty to improve efficiency. A successful example is the GECAM-HXMT-SVOM (\textit{GHS}) constellation collaboration, which sets a precedent for other $\gamma$-ray monitor constellations. 
However, joint duty also presents challenges to Burst Advocates (BAs), including the increased number of triggers and, more importantly, the frequent switching between various systems due to incompatibilities among different missions, which complicates the situation. 
To address the current requirements of multi-wavelength and multi-messenger astronomy, we developed a customized framework for unified trigger processing within the \textit{GHS} joint duty, named ``BA’s Rapid Evaluation and Analysis Kit for Formulating Alerts and Summary Tools" (\texttt{BREAKFAST}). 
This framework incorporates a series of automated, semi-automated, and manual pipelines designed to rapidly process triggers of prompt emissions in the gamma-ray band from different instruments, while maintaining flexible compatibility for future missions. 
The pursuit of \texttt{BREAKFAST} goes beyond merely providing trigger processing for BAs. \texttt{BREAKFAST} also aims to filtering high-value targets and guiding follow-up telescopes through rapid analysis and reporting, thus serving as an important bridge between prompt emission observations and afterglow observations. To this end, a suite of comprehensive analysis modules is included in \texttt{BREAKFAST}, particularly the specially designed module that predicts X-ray afterglow brightness based on prompt emission properties. 
The framework’s effectiveness has already been demonstrated in recent observational campaigns, and it is expected to play a significant role in the discovery and observation of peculiar transients in the future.
\end{abstract}

\section{Introduction} 
Time-domain and multi-messenger astronomy (TDAMM), which focuses on exploring transient phenomena across multi-wavelength and multi-messenger, plays a crucial role in the study of astrophysics and cosmology. As the primary research objects in TDAMM, high energy transients such as Gamma-ray Bursts (GRBs) \citep[see e.g.,][and references therein]{GRB_1,GRB_2,GRB_3} and Magnetar X-ray Bursts (MXBs) \citep[see e.g.,][and references therein]{MXB_1,MXB_2,MXB_3,MXB_4} are mainly  first detected and observed by wide Field-of-View (FoV) $\gamma$-ray monitors, including Gravitational Wave High-energy Electromagnetic Counterpart All-sky Monitor (GECAM) \citep{GEC_INS_Li2022,GECAM_review}, \textit{Fermi} Gamma-ray Burst Monitor (\textit{Fermi}/GBM) \citep{GBM_meegan_09}, Hard X-ray Modulation Telescope (\textit{Insight}-HXMT) \citep{HXMT_zhang_2020}, \textit{Swift} Burst Alert Telescope (\textit{Swift}/BAT) \citep{Swift_1,Swift_2}, Space-based multiband astronomical Variable Objects Monitor Gamma Ray burst Monitor (SVOM/GRM) \citep{SVOM_1,SVM_GRM_Dong2010} and so on, which are then followed by joint observations at other wavelengths such as soft X-ray (e.g., Einstein Probe (EP) \citep{EP_1} , \textit{Swift}/XRT \citep{Swift_3}, et al.), optical (e.g., \textit{Swift}/UVOT \citep{Swift_4}, SVOM/VT\citep{VT_1}, et al.). 

Usually each $\gamma$-ray monitor collaboration has a team of Burst Advocates (BAs) (or Transient Advocates, dubbed as TAs, playing similar roles) that take personal responsibility for an individual burst, monitoring the trigger state and conducting preliminary data reductions to support the community in performing and coordinating joint observations. 
However, with the rapid development of observations and theories, as well as the upgrades and additions of instruments, the paradigm of observational campaign has also been changing drastically, which places much more requirements on BA's work than before. 

On one hand, the observation campaigns are still usually initiated by $\gamma$-ray monitors, whose number has grown rapidly. Therefore, it is essential to promptly report trigger information and preliminary analysis results (e.g. by GCN Circular) to assist the community in selecting high-value targets for follow-up observations. Additionally, due to the usually poor localization accuracy of single $\gamma$-ray monitors independently, it has become increasingly important to quickly generate IPN format data for multi-instrument joint localization. 
On the other hand, a growing number of wide FoV multi-wavelength telescopes (e.g. EP/WXT \citep{EP_1} and SVOM/ECLAIRs \citep{ECLAIRs} in X-ray and hard X-ray, CHIME \citep{CHIME} in radio, Legacy Survey of Space and Time (LSST) \citep{LSST} in optical) and multi-messenger instruments (e.g. LIGO-Virgo-KAGRA (LVK) in Gravitational Wave (GW) \citep{170817A_1,170817A_2}, Super-Kamiokande \citep{Super_K} and IceCube \citep{ICECUBE} in neutrino) are also increasingly becoming initiators of joint observation campaigns in TDAMM. In this case, it is necessary to examine the visibility of the $\gamma$-ray monitors and conduct a hierarchical signal search (i.e. blind search and target search), then the upper limits are need to be reported if no $\gamma$-ray counterpart is identified. 
Hence, BAs have to transform from being merely the initiator of the follow-up joint observations in the past, which typically just focused on the trigger status of their own $\gamma$-ray monitors, to a responder of transients detected by instruments of other wavelengths and messengers, constantly needing to pay attention to the trigger information of other instruments. 

With the communication and developing of the TDAMM community, many researchers are involved in the BA duty of multiple missions. 
An example is the GECAM-HXMT-SVOM (\textit{GHS}) constellation collaboration, which consists of as many as six instruments. 
The synergy among the GECAM, HXMT, and SVOM/GRM science team, as well as operation teams, has led to the possibility of integrating their BA teams for these instruments into a single entity, meaning that the BA is now responsible for processing triggers from all these instruments simultaneously. 

Such joint BA duty significantly enhanced the efficiency and response speed. 
The mutual compensation of multi-instruments' FoV also significantly enhances the sky coverage and monitoring time coverage, which is beneficial to TDAMM joint observation campaign. 
However, the increase in analysis tasks and the differences between the analysis process of these instruments have made it difficult to fully realize the advantages of these joint BA duties. 
Hence a well-designed data reduction framework is also needed to assist the BA in performing their duties. 
Thanks to the far-sighted multi-instrument pipeline \texttt{ETJASMIN} (Energetic Transients Joint Analysis System for Multi-INstrument, \citep{ETJASMIN_I,ETJASMIN_II}), which was originally designed for GECAM, a substantial amount of preliminary software has been developed, as well as the data reduction software \texttt{GECAMTools}\footnote{Originally designed for GECAM data extraction, and has now been extended to data extraction for HXMT and SVOM/GRM.} \citep{GECAMTools} and spectral fitting software \texttt{ELISA} \citep{elisa2024}. 
It has become feasible to conduct additional design to develop a framework tailored to the characteristics of the BA duty of \textit{GHS}.

Here we introduce the pipeline framework of ``BA’s Rapid Evaluation and Analysis Kit for Formulating Alerts and Summary Tools", abbreviated as \texttt{BREAKFAST}. 
This framework is the first one designed for joint observations and duties of $\gamma$-ray constellation across multiple instrument collaboration, aimed at assisting BAs in summarizing trigger information, conducting data reduction, and generating necessary analysis results for follow-up community (e.g. post GCN circulars) by a hierarchical system with automatic, semi-automatic and manual pipelines. 
Considering that each instrument has developed and deployed different pipelines adapted to their own data and observation strategies, the \texttt{BREAKFAST} does not aim to reconstructing or completely replacing all the previous pipelines, but rather attempts to integrate them through a top-level framework design in a way that invokes their functionalities, thereby preserving sufficient flexibility and reducing development efforts. 

This paper is organized as follows. 
The instrument and the data that are involved in \texttt{BREAKFAST} are described in Section \label{sec:instrument}. 
Section \label{sec:design} presents the design conception and architecture of the framework. 
Then some cases of observations based on the \texttt{BREAKFAST} framework are discussed in \label{sec:case}
Finally,we give a discussion and conclusion in \label{sec:summary}.

\section{Instruments and Data} \label{sec:instrument}
For the \texttt{BREAKFAST} framework that targets multiple instruments, it is necessary to design compatibility by integrating the characteristics and data types of different instruments. 
Based on data access permissions, we categorized these instruments involved in\texttt{BREAKFAST} as internal instruments and external instruments. 
On the other hand, the instrument should also classified as ``High precision localization capacity" (e.g. \textit{Swift}/BAT, \textit{Fermi}/LAT, SVOM/ECLAIRs, et al.), ``good localization capacity" (e.g. \textit{Fermi}/GBM, GECAM, et al.) and ``poor/no localization capacity" (e.g. \textit{Insight}/HXMT, SVOM/GRM, INTEGRAL/SPI-ACS). We need to note that such distinctions are not strict, as data from instruments like \textit{Fermi}/GBM is publicly available and can therefore be directly used for analysis, and data from SVOM/ECLAIRs and EP are not accessible, due to the data policies, for this framework now, although some BAs of \textit{GHS} also participate in their duties. 

\subsection{Internal instrument}
The internal instruments are the prime object that \texttt{BREAKFAST} is designed for, currently mainly including the series of satellites from GECAM constellation, \textit{Insight}-HXMT, and SVOM, that is, the \textit{GHS} mentioned earlier. 
In fact, although there is already six instruments incorporated in the joint BA duty and \texttt{BREAKFAST}, the instrument list is not fixed, and there is still potential for expansion in the future; new instruments such as POLAR-2 and GRID constellation can also be added flexibly in the future. 

\subsubsection{GECAM}
GECAM is a satellite constellation dedicated to monitoring all-sky high-energy astrophysical transients as well as terrestrial transient phenomena. At the time of writing, GECAM mission comprises four instruments: GECAM-A, GECAM-B \citep{GEC_INS_Li2022}, GECAM-C \citep{HEBS_INS_Zhang2023, 2024NIMPA105969009Z}, and GECAM-D \citep{GTM_INS_wang2024,GTM_INS_Feng2024}. As a versatile instrument, GECAM has achieved a series of discoveries in high-energy transient phenomena \citep{GECAM_review}, including gamma-ray bursts (GRBs) \citep{GECAM_GRB_1,GECAM_GRB_2,GECAM_GRB_3,GRB240825A,07A_Yi}, soft gamma-ray repeaters (SGRs) \citep{GECAM_SGR_1,GECAM_SGR_2}, high-energy counterparts of Gravitational Wave (GW) and Fast Radio Burst (FRB) \citep{GECAM_221014_ATEL}, Solar Flares (SFLs) \citep{GECAM_SF}, as well as Terrestrial Gamma-ray Flashes (TGFs), Terrestrial Electron Beams (TEBs) \citep{GECAM_TEB} and peculiar oscillating particle precipitation events \citep{OPP_wang}. 

GECAM-A and GECAM-B were launched on December 10, 2020 (Beijing time) into a Low Earth Orbit (LEO) with an altitude of about 600\,km and an inclination angle of 29 degrees. Each satellite is equipped with two types of scintillator-based detectors: gamma-ray detectors (GRDs) and charged particle detectors (CPDs) \citep{GEC_INS_An2022,GEC_INS_Xv2021,CPD_calibration}. Both GECAM-A and GECAM-B are equipped with 25 GRD detectors and 8 CPD detectors. 
GECAM-C is one of the payloads onboard the SATech-01 satellite, which was launched on July 27, 2021 and quit operation on February 13, 2025. SATech-01 operated in a Sun-Synchronous Orbit (SSO) with an inclination angle of 97.4$^\circ$ and an altitude of about 500\,km. GECAM-C comprises two hemispherical domes: one mounted on the top and the other on the bottom of the satellite platform. Each dome hosts six GRDs and one CPD. This configuration provides GECAM-C with a nearly 4$\pi$ (full-sky) field of view. 
GECAM-D is the 4th instrument in GECAM series which was launched on 13 March 2024 on board the DRO-A satellite. 

GECAM features a variety of well-designed data types with different resolution and time delay. 
First of all, GECAM has onboard trigger capabilities and innovatively utilizes BeiDou navigation system to quickly downlink trigger alerts in the form of BeiDou short messages (BDM) with a time delay of $\sim$minutes \citep{GECAM_trigger,GECAM_alert}. The BDM data includes a series of lightcurves and spectra, as well as in-flight localization obtained by counts rate moduation with fixed template. 
With the downlinked BDM data in-flight trigger data, a more accurate on-ground localization by joint fitting of localization and spectrum will be performed \citep{GECAM_loc_liao,YIZ_LOC_GEC_2023}. 
Besides the BDM data of in-flight trigger, a high sensitivity blind search is automatically conducted for the event-by-event (EVT) data and binned time data (BTIME) \citep{GECAM_ground_search}, which usually have a time delay of a few hours to $\sim$day, and on-ground localization can also be obtained automatically \citep{GECAM_loc_liao,YIZ_LOC_GEC_2023}. 

Although the GECAM data directly used in \texttt{BREAKFAST} is primarily the continuous Level 1 daily EVT data, other types of GECAM data are also indirectly used by invoking the results of other pipelines (e.g. the on-ground search, the trigger classification and so on).

\subsubsection{\textit{Insight}-HXMT}
\textit{Insight}-HXMT (or HXMT) is China's first X-ray astronomy satellite, which was proposed in 1993 and launched on June 15, 2017, operating in a LEO orbit with an altitude of $\sim$550\,km and inclination of $\sim43 ^\circ$. 
HXMT consists of three collimated telescopes, namely the high energy X-ray telescope (HE), the medium energy X-ray telescope (ME) and the low energy X-ray telescope (LE) \citep{HXMT_zhang_2018,HXMT_zhang_2020,HXMT_Li_2020}. HE consists of 18 NaI(Tl)/CsI(Na) phoswich scintillation detectors with collimators. For GRB detection, gamma-ray can penetrate the satellite platform and be detected by CsI detectors of HE with a large effective area \citep{HE_calibration,HE_calibration_updated}. 
HXMT does not equipped with in-flight trigger function. Therefore, only on-ground search and on-ground triggers are enabled for continuous Level 1K daily EVT data, which is also utilized in \texttt{BREAKFAST}. 
Additionally, HXMT also cannot independently localize GRBs unless the transient happens to fall within the field of view of the collimator.

\subsubsection{SVOM}
\textit{SVOM} (Space-based multiband astronomical Variable Objects Monitor, \cite{SVOM_1}) is a Sino-French mission primarily for GRB studies, aiming to the discovery and multi-wavelength follow-up of cosmic transients. 
The mission includes a satellite, which was launched on 22 June 2024, and a series of dedicated ground telescopes. 

The satellite carries two wide-field monitors for GRB prompt emission detection, called ECLAIRs (coded-mask gamma-ray imager, operating in energy between 4 and 250 keV) and GRM (Gamma-Ray Monitor). ECLAIRs is in charge of providing highly precise localization and soft X-ray energy coverage, while GRM is responsible for obtaining the wide energy spectrum measurement, as well as a wide FOV monitor. 
Besides these two monitors, two narrow-field telescopes for GRB follow-up are also equipped, called MXT (Microchannel X-ray Telescope, energy between 0.2 and 10 KeV) and VT (Visible Telescope, from 450 to 650 nm for the blue channel and from 650 to 1000 nm for the red channel). 
The on-ground including two sets of ground wide-angle camera arrays (GWAC) and two ground follow-up observation telescopes, named as Chinese Ground Follow-up Telescope (C-GFT) and French Ground Follow-up Telescope (F-GFT). 

The SVOM payloads directly included in the current \textit{GHS} joint duty are mainly SVOM/GRM, although SVOM/ECLAIRs, SVOM/VT, and GWAC also play important roles in the \texttt{BREAKFAST} framework of \textit{GHS}, which will be discussed in more details in Section\,\ref{sec:design}. The main sensitive detector of GRM consists of three wide field of view NaI(Tl) scintillation Gamma-Ray Detectors (GRDs), which are labeled as GRD01, GRD02 and GRD03, all operating in the 15 keV to 5 MeV energy range \cite{SVM_GRM_Dong2010}. 
All GRM GRDs operate in two readout channels: high gain (HG) and low gain (LG), and only if the deposited energy is higher than the energy range of HG channel, the event will be processed by the LG channel. 

Similar to GECAM, GRM is also equipped with real-time triggering and localization capabilities \citep{GRM_trigger}. 
The triggering of GRM is divided into model-1 and model-2 based on the presence of ECLAIRs triggers. 
Depending on whether ECLAIRs is triggered or not, the GRM in-flight triggers have two types: mode-1 (GRM-only trigger) and mode-2 (ECLAIRs\&GRM trigger). 
Once in-flight triggered, the trigger information, which mainly includes three sets of lightcurves with different time resolutions, can be downlinked by the VHF system with low latency. 

In addition to VHF data, GRM also has continuous EVT data (untilized in \texttt{BREAKFAST}) and the binned time$\&$energy (spechist) data.
Whenever new continuous daily data (Level 1B) is received, the automatic on-ground blind search, which is migrated from GECAM on-ground search pipeline, will be initiated.

Although GRM can also independently localize transients, the in-flight trigger is not always able to provide accurate localization due to limitations in onboard software. Moreover, reliable localization can only be obtained through joint fitting of localization and spectrum within a certain field of view range due to the limited number of probes in GRM. 
Hence GRM is also classify as ``poor/no localization capacity" instrument in \texttt{BREAKFAST}, and thus only bursts that are joint detected by other ``good localization capacity" instruments can have their spectra analyzed with GRM data.

\subsection{External instrument}

Generally speaking, the external instrument includes all instruments except \textit{GHS}. 
Here we only discuss the instruments directly associated with \texttt{BREAKFAST}. 
Data from other external instruments, apart from \textit{Fermi}/GBM, are not directly used in \texttt{BREAKFAST}, but serving as provider of trigger time and localization. However, we need to note that all instruments, whether discussed here or not, play important roles in the TDAMM joint observation and may also have direct interaction with \textit{GHS} in future joint operations.

\subsubsection{Fermi}

The Gamma-ray Burst Monitor (GBM) is one of the two instruments onboard the \textit{Fermi} Gamma-ray Space Telescope \citep{GBM_meegan_09,GBM_Bissaldi_09}. \textit{Fermi}/GBM is composed of 14 detectors with different orientations: 12 Sodium Iodide (NaI) detectors (labeled from n0 to nb).  For spectral analysis, the energy range is 8-900\,keV for NaI detectors and 0.3-40\,MeV for BGO detectors. 
Thanks to the publicly available policy of \textit{Fermi}/GBM, many flexible pipelines for GBM data analysis and research have been developed by TDAMM community, for example: \texttt{RapidGBM} \citep{RAPIDGBM}. 
As one of the most important references for \textit{GHS} joint duty and external trigger origin, \textit{Fermi}/GBM data are also integrated into \texttt{BREAKFAST}. 
The first type of GBM data used in \texttt{BREAKFAST} is the TRIGDAT, which is the burst alert telemetry, and can be obtained with low latency, providing lightcurve, spectra and localization. 
The other type of GBM data used in \texttt{BREAKFAST} is the Time-Tagged Events (TTE), which is treated the same as the internal instrument. 

The Large Area Telescope (LAT) is the other instrument onboard the \textit{Fermi} Gamma-ray Space Telescope, which is a pair-conversion telescope comprising a 4 × 4 array of silicon strip trackers and cesium iodide (CsI) calorimeters \citep{LAT_Atwood_09}. 
LAT data are not directly input to \texttt{BREAKFAST} but are only used as an instrument with "high precision localization capacity" to update localization, which is obtained by GCN Notice.

\subsubsection{\textit{Swift}, Konus-Wind and other IPN instruments}
Two instruments onboard \textit{Swift} was used in \texttt{BREAKFAST}: Burst Alert Telescope (BAT) and X-Ray Telescope (XRT). 
The former is used to provide real-time external triggering and high precision localization, and the information are received through GCN Notice. 
The archived X-ray afterglow lightcurves of the latter \citep{xrt_catalog_1,xrt_catalog_2} is used to estimate the X-ray afterglow brightness and redshift of newly detected GRBs. 

As for Konus-Wind and other IPN instruments, on one hand, they serve as one of the sources for external triggers, and will be added into the trigger database manually. 
On the other hand, IPN localization will be used to manually update localization results. 
Additionally, one of the data products from \texttt{BREAKFAST} is the IPN format light curve, which will also be provided to the Konus-Wind team for IPN triangulation.

\section{Design and Architecture of \texttt{BREAKFAST}} \label{sec:design}

\begin{figure*}
\centering
\begin{overpic}[width=0.9\textwidth]{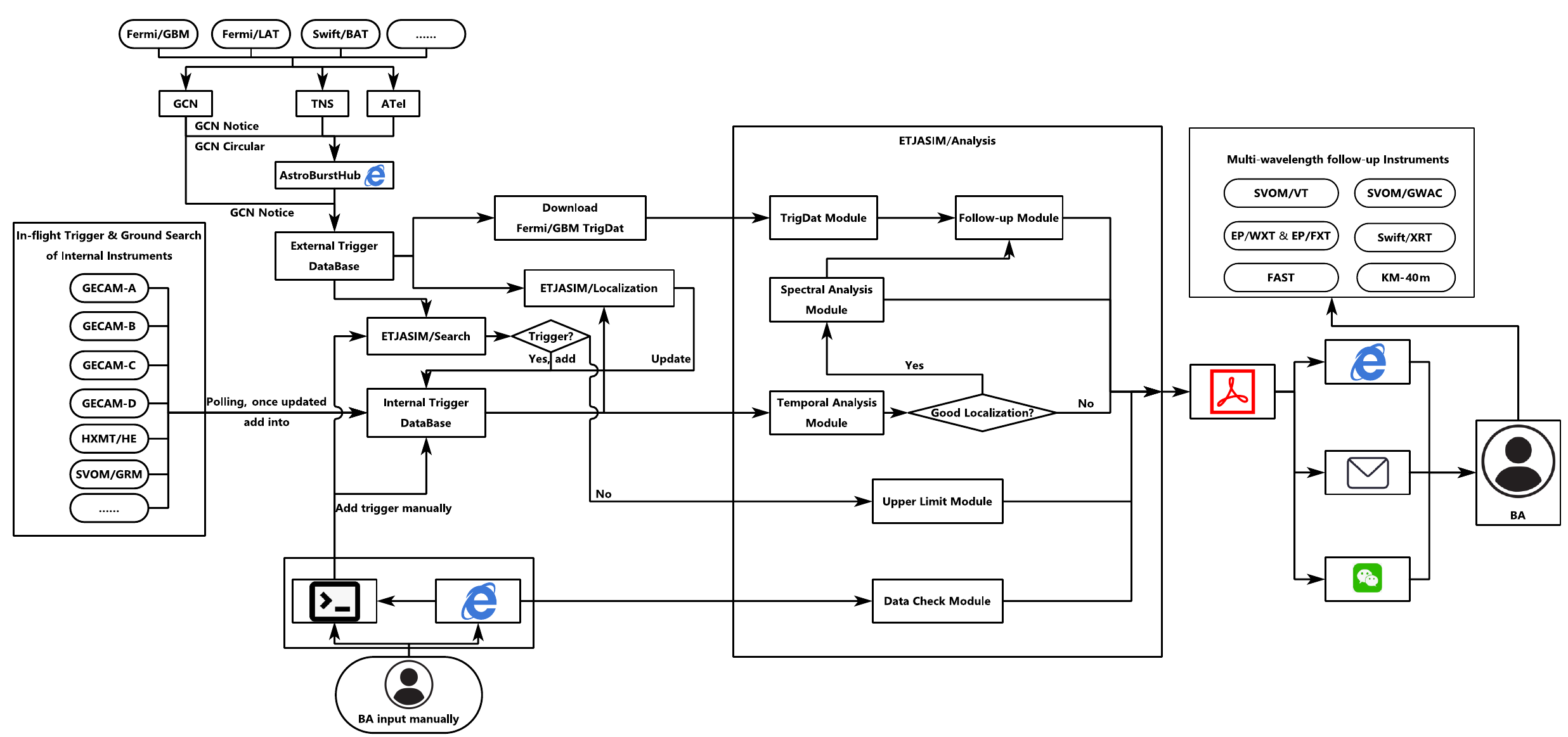}\end{overpic} 
\caption{\textbf{The architecture of \texttt{BREAKFAST} framework.} 
The new received triggers, including both internal triggers and external triggers, will waken the automatic analysis flow. 
If it is a new trigger (with no time coincidence with any previous triggers), the automatic analysis will be directly performed on all internal instrument data that has already been received. 
While if there is a time coincidence with past triggers (in which case, the triggers that match in time will be automatically merged into one burst), it will first be determined whether the burst information needs to be updated. If the burst information is updated, the automatic analysis will be re-conducted. 
For semi-automatic analysis flow and manual analysis flow, the trigger information can be input via terminal or webpage interface. 
The analysis result will be output and combined as a PDF file, which will be posted to BA by E-mail or checked via a webpage. 
The follow-up observation will be requested by BA after evaluating the importance with the analysis result provided by \texttt{BREAKFAST}.}
\label{fig:Fig_arch}
\end{figure*}

The design concept and architecture of \texttt{BREAKFAST} is introduced in more details in this section. 
The flowchart of \texttt{BREAKFAST} is depicted in Figure\,\ref{fig:Fig_arch}. 
According to the different roles played by each pipeline, this entire framework can be roughly divided into three general modules: the Trigger module, the Analysis module, and the Interactive module. We need to acknowledge that \texttt{BREAKFAST} is still undergoing constant iteration and updates, therefore, although some basic and fixed designs are introduced here, adjustments and modifications may still be possible in the future to ensure the best functionality of \texttt{BREAKFAST}.

\subsection{Trigger Module}
The first primary part of \texttt{BREAKFAST} is the trigger module. 
On one hand, thanks to the good design and high performance of the search system, no search is conducted again in the automatic flow of \texttt{BREAKFAST}. \texttt{BREAKFAST} will poll the internal server (data center in IHEP, which achieve centralized \textit{GHS} science data and engineering data) to search for new internal triggers (including in-flight and on-ground). 
On the other hand, \texttt{BREAKFAST} can receive GCN Notices through \texttt{Kafka} and parse them to obtain the trigger time and localization of external triggers. 

However, some transients are reported in the form of natural language (e.g., GCN Circular and ATel), whose information is difficult to be extracted automatically because the expression habits are not completely consistent among different teams. 
To deal with this, \texttt{BREAKFAST} leverages another platform, Astronomical Burst Hub (abbreviated as AstroBurstHub) \footnote{http://www.astrobursthub.cn/}, which is also developed by the \textit{GHS} team and powered by a large language model (LLM), to receive and parse the information of these transients from various platforms that collect and distribute messages publicly, and then these information will be added to \texttt{BREAKFAST} as external triggers. 

Both new internal triggers and external triggers will be written into a \texttt{SQLite}-based database. 
The internal trigger database mainly includes the following parameters:
\begin{enumerate}[(1)]
\item trigger UTC time, RA, DEC: the basic trigger time and localization of a trigger.
\item trigger time origin, localization origin: the instrument that provides trigger time and localization. 
\item auto-type, manual-type: the trigger classification by automatic and manual, including ``GRBs", ``SGRs", ``Solar Flares", ``Charge particles" and ``others". 
\item the observation state of each internal instrument, which need to be selected by BA manually after checked the results, including ``true burst", ``burst-candidate", ``no data", ``no signal", ``false trigger". 
\item the process state of each internal instrument, including ``waiting", ``running" and ``finished".
\item the waiting time since the trigger time of each internal instrument: if the waiting time for a certain instrument exceeds 3 days, it will be deemed as not collecting data at that time, and the process state will be set to ``finished", meaning it will no longer be automatically submitted for analysis. 
\end{enumerate}

A basic design of \texttt{BREAKFAST} is to process all available data (i.e., the data received) for each trigger (including both internal triggers and external triggers) received, specifically for that trigger time and localization (if available). 
Triggers from different instruments with the same trigger time (accurate to the minute) will be archived together and marked as ``burst-candidate" automatically (can be changed manually by BA) for all triggered instruments, as a burst activity of the celestial object may be observed by one or multiple instruments and lead to more than one triggers with similar trigger time, therefore, triggers occurring at close intervals may indicate that they originate from the same burst process. 
One of the trigger times will be selected as the unified trigger time of the burst for subsequent analysis of all the instruments. 
The automatically selected unified trigger time follows the following logic:
\begin{enumerate}[(1)]
\item Prioritize the selection of internal trigger time than external trigger time.
\item For internal triggers, prioritize the selection of in-flight trigger time than on-ground trigger trigger. The priority order of the instruments is GECAM-B, GECAM-C, SVOM/GRM, HXMT, GECAM-A, and GECAM-D.
\item For external triggers, prioritize the selection of \textit{Swift}/BAT trigger time than \textit{Fermi}/GBM and other instruments.
\end{enumerate}
For localization, a similar situation exists. The automatic flow will be initialed once the localization is updated. The logic for automatically selecting localization is
\begin{enumerate}[(1)]
\item Prioritize the selection of on-ground localization than in-flight localization, and prioritize the selection of localization provided by ``high precision localization capacity" instruments.
\item For localization provided by ``good localization capacity", the priority order of the instruments is GECAM-B, GECAM-C, \textit{Fermi/GBM}, GECAM-A. While GECAM-D, SVOM/GRM and HXMT are considered as ``poor/no localization capacity" instrument.
\item For localization provided by ``good localization capacity", prioritize the selection of \textit{Swift}/BAT localization than \textit{Fermi}/LAT and other instruments.
\end{enumerate}

In addition, triggers can also be manually added to the trigger database.

\subsection{Analysis Module}
\begin{figure*}
\centering
\begin{overpic}[width=0.9\textwidth]{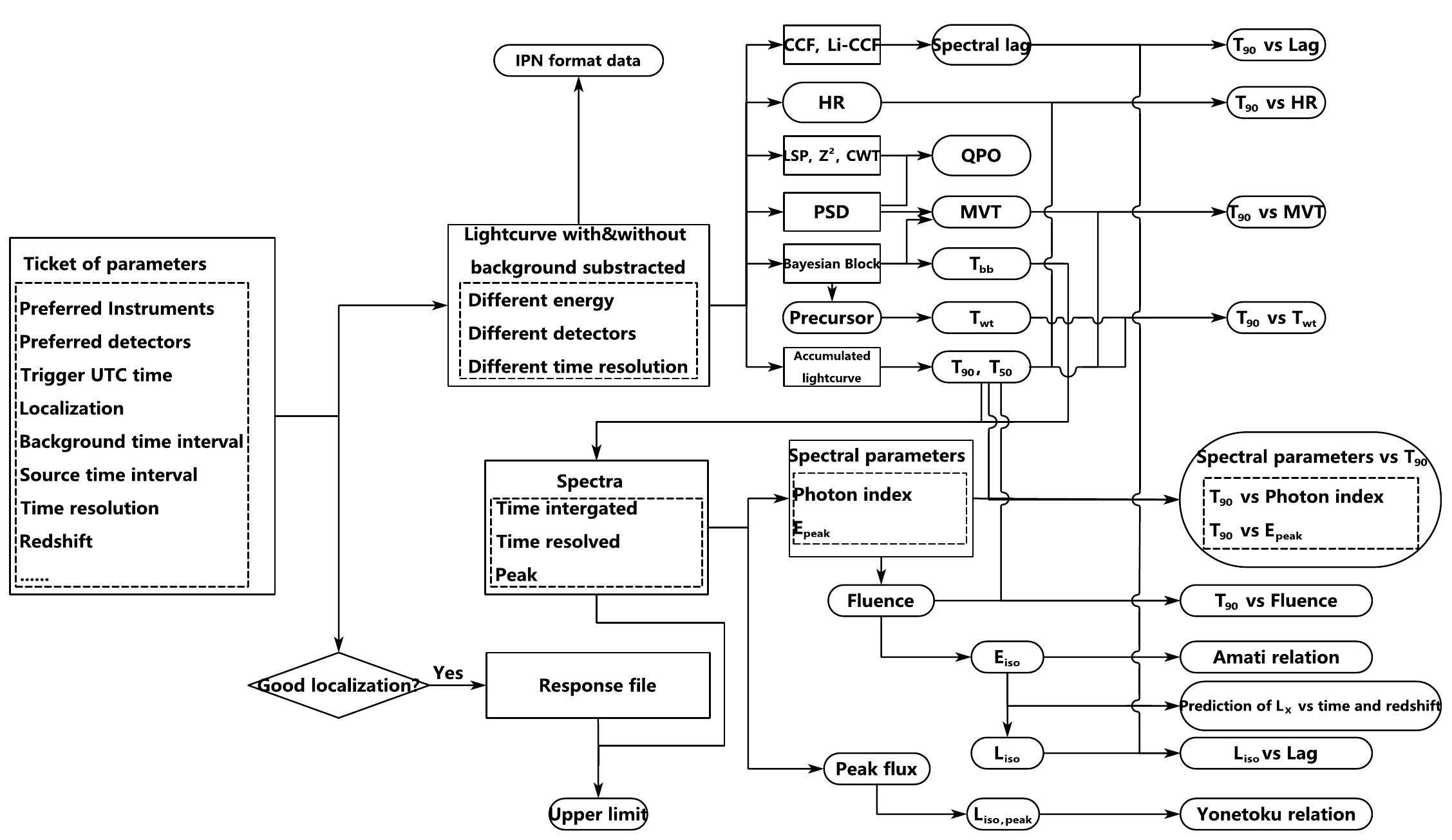}\end{overpic} 
\caption{\textbf{The analysis for triggers includes in \texttt{BREAKFAST}.} 
The rectangle represents some analysis process or basic data products in a unified format for multiple instruments, while the rounded rectangle represents the analysis results that can be used for preliminary science discussion.}
\label{fig:Fig_ana}
\end{figure*}

The second primary part of \texttt{BREAKFAST} is the analysis module, which is included the \texttt{ETJASMIN}/Analysis part marked in Figure\,\ref{fig:Fig_arch}. 
A more detailed flowchart of the analysis module is separately shown as Figure\,\ref{fig:Fig_ana}.

\subsubsection{Single Instrument}
Considering the different data extraction tools for each instrument and its data, the part for generating lightcurves and spectra in \texttt{BREAKFAST} are not shared among different instruments and data types. 
While the extracted lightcurve and spectra are saved with the same format, allowing subsequent temporal and spectral analyses to share the same pipeline. 

\begin{figure*}
\centering
\begin{overpic}[width=0.8\textwidth]{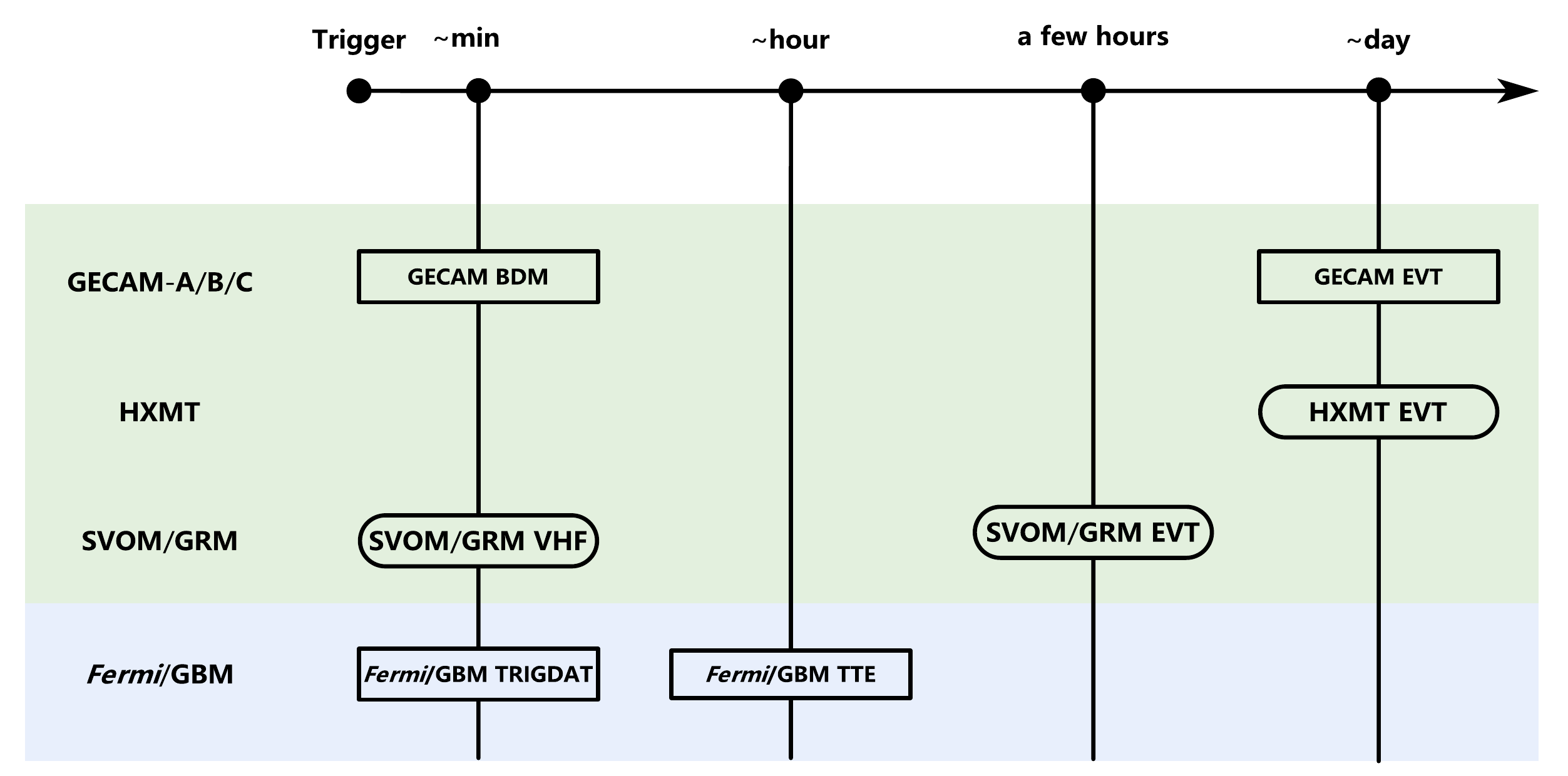}\end{overpic} 
\caption{\textbf{The instruments with their various types of data utilized in \texttt{BREAKFAST}}. 
The time delay from trigger time to being available for scientific analysis for each type of data is also labeled roughly. 
The green part includes the internal instruments and the blue part includes the external instruments. 
The rectangle represents the data type with good localization capacity, while the rounded rectangle represents the data type with poor or even no localization capacity.}
\label{fig:Fig_data}
\end{figure*}

The automatic flow begins with temporal analysis, including the calculation of $T_{90}$, spectral lag, hardness ratio, and so on, as portrayed in Figure\,\ref{fig:Fig_data}. 

Due to the fact that localization information is not always available, sometimes the automatic flow will stop before the spectral analysis begins. 
Three sets of spectra with different time ranges are prepared for the spectral analysis, including a time integrated spectrum, a peak time spectrum and a series of time resolved spectra, and the latter two only apply to GRBs with $T_{90}$ greater than 2 seconds. 
For the time integrated spectrum, the time interval is set as the $\rm max\{{T_{bb},T_{90}}\}$. 
For the peak time spectrum, the time interval is set as the time segment with maximum counts within 1 second. 
 nFor the time resolved spectra, a series of spectra is produced every 1 second from the start time to the end time of $T_{90}$. 
The fitting is performed with the \texttt{ELISA} package, with three models: Powerlaw (PL), Cutoff powerlaw (CPL) and BAND model, while only the result of CPL will be used for subsequent analysis in the automatic flow. 

Once the spectral analysis is completed, a series of binary variable comparisons will be automatically performed subsequently, including Amati relation, Yonetoku relation and so on. 
All the generated images and results will be directly used for writing the GCN Circular. 

The above process is mainly aimed at EVT data. 
In fact, \texttt{BREAKFAST} is also specially designed with modules to analyze and process real-time trigger data, including BDM data of GECAM, TRIGDAT of \textit{Fermi}/GBM, and VHF data of SVOM/GRM, with similar process. 
Among them, the BDM data of GECAM and the TRIGDAT of \textit{Fermi}/GBM can be used for both temporal and spectral analysis, while the VHF data of SVOM/GRM can only be used for temporal analysis. 

Additionally, there are two specially added analyzes in \texttt{BREAKFAST}. The first is the identification of type IL GRBs, as one of the scientific objectives of SVOM is the observation of kilonovae. 
An analysis of precursors (e.g. $T_{90}$-$T_{\rm wt}$) is added in the temporal analysis, with the same method of \citep{Type_IL_2} to searching for possible burst that follow the Type IL burst pattern. 
When $T_{90}$-$T_{\rm wt}$ diagram shows that a burst has a long $T_{90}$ and a short $T_{\rm wt}$, and is consistent to the type I region in the Amati relation diagram at any redshift. 
A request for follow-up observation by SVOM/GWAC, which has a large FoV, will be sent. If the localization error box is small enough, follow-up observation by EP/FXT and SVOM/VT will also be requested. 

The other analysis is to predict the X-ray afterglow brightness based on the brightness of $\gamma$-ray prompt emission. 
The context of this analysis is that the high sensitivity and large Fov of EP/WXT make it possible to search for X-ray afterglows with EP/WXT. 
The observation of EP/WXT can significantly reduce the localization error box obtained from $\gamma$-ray monitor, which is usually several degrees for $\gamma$-ray monitors and only several arcmins for EP/WXT. 
The prediction is mainly inspired by the closure relation of $L_{\rm X}/E_{\rm iso}$ and the rest-frame time \citep{pre_ag_1,pre_ag_2}, where $L_{\rm X}$ is the 2-10\,keV X-ray luminosity in the rest frame and $E_{\rm iso}$ is the prompt emission isotropic energy. 
By fitting the X-ray afterglow dataset from \cite{pre_ag_1} with a simple powerlaw function, we can obtain a relation of $L_{\rm X}/E_{\rm iso}\propto t_{\rm rest}^{-1.23}$, which is to predict the X-ray afterglow flux with an assumption of redshift at some fixed values.

\subsubsection{Multiple Instruments}
The multi-instrument analysis in \texttt{BREAKFAST} is mainly based on \texttt{ETJASMIN}/Search and \texttt{ETJASMIN}/Localization. 
A much more sensitive target search pipeline (i.e. \texttt{ETJASMIN}/search \citep{ETJASMIN_II}) can be invoked in the semi-automatic flow and manual flow. 
Moreover, if a burst is detected by more than one internal instruments, a joint localization based on time delay with advanced Li-CCF \citep{joint_localization_1} can also be conducted with the ETJASMIN/localization pipeline \citep{ETJASMIN_I}. 
Although the function have been developed, there is no joint spectrum fitting in both the automatic and semi-automatic flow in \texttt{BREAKFAST}, as this is far beyond the requirement of preliminary analysis for BA duty.

In addition, there is a separate flow that can bypass other analysis processes (i.e. add to trigger database and preform temporal analysis) to generate information such as all-instruments lightcurves and nadirs for the BA to quickly check the data, which is also prepared for instrument scientists and operation team.

\subsection{Interactive Module}
To accommodate various situations as much as possible and based on practical feedback from BA operations, \texttt{BREAKFAST} has designed two main interactive methods: an interactive script in the terminal and a lightweight web interface based on \texttt{flask}; 
although the interactive methods for the automatic flow, semi-automatic flow, and manual flow have slight differences. For the automatic flow, no input from a BA is required. 
A BA only needs to check the various output results on the website and confirm the classification of each trigger. 
In addition, the automatic process results will also be sent to all BA members via email and WeChat. For the semi-automatic flow, a sheet with limited modifiable parameters need to be filled out through the terminal or the website, and submitted to the trigger database. Then the process will be the same with the automatic flow, but with updated parameters. 

To simplify operations as much as possible, even in the semi-automatic flow, the parameters that can be modified are limited, and more parameters still use default values that are not visible in the front end. 
Typically, the automatic flow and semi-automatic flow are sufficient to solve the vast majority of works for a BA. 
However, there are always some cases that require modifications to these typically invisible parameters, and the manual flow has been retained precisely for this purpose. 
In this scenario, a BA needs to access the backend to directly modify the parameter list file and then submit it. 

It is also possible to run only certain modules via the interface if the user is familiar with this framework. 
For example, ETJASMIN/Search can be directly invoked from the terminal for target search. 
The function to calculate the upper limit can also be directly invoked without the need to initialize the search or add triggers to the trigger database. 
This brings a relatively high degree of freedom and flexibility, but it also inevitably leads to an increase in operational complexity, and not included in the normal flow in Figure\,\ref{fig:Fig_arch} and Figure\,\ref{fig:Fig_ana}. 
Therefore, the main operating modes of BREAKFAST remain the automatic flow and semi-automatic flow.

\section{Case study} \label{sec:case}
\texttt{BREAKFAST} has been undergoing test in the \textit{GHS} joint duty for a period of time and plays an important role in many time of TDAMM joint observation, including the observation of peculiar GRBs and magnetar X-ray bursts. 

\subsection{Bright GRBs and afterglow prediction}
\begin{figure*}
\centering
\begin{tabular}{cc}
\begin{overpic}[width=0.45\textwidth]{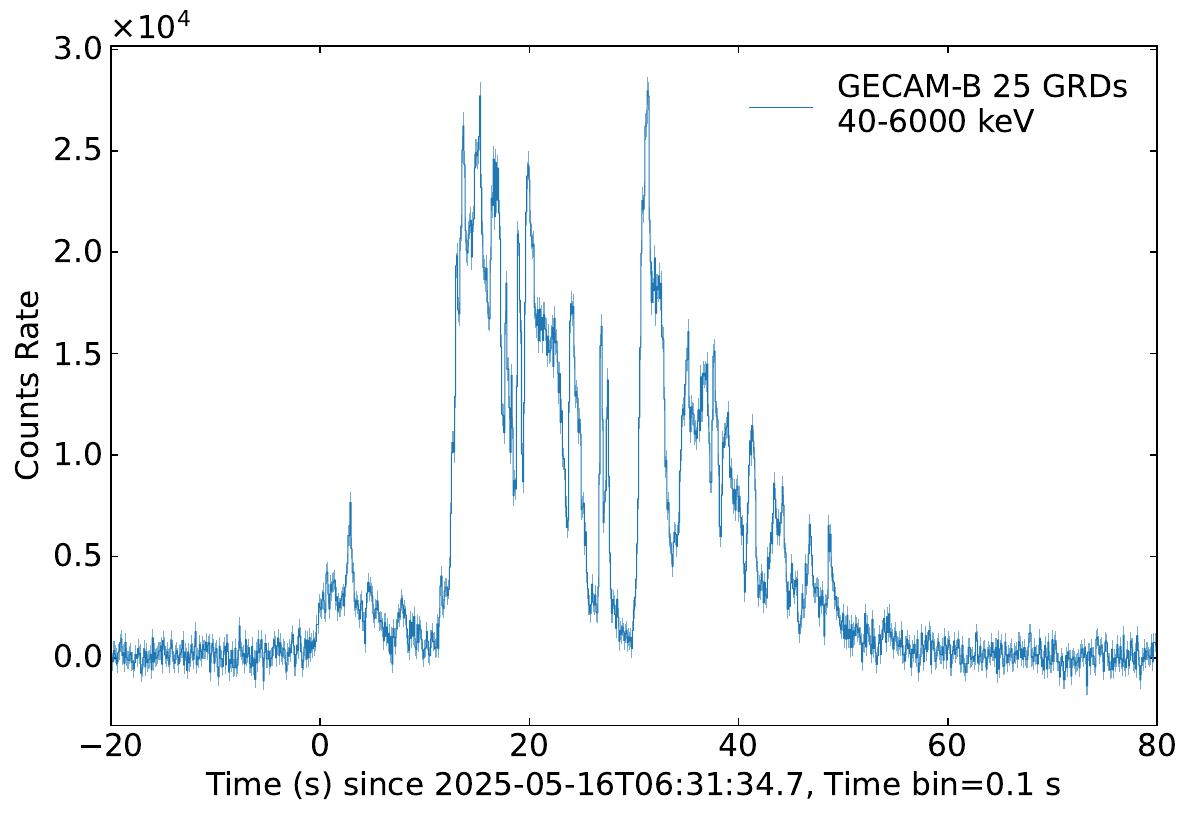}\put(-3, 65){\bf a}\end{overpic} &
        \begin{overpic}[width=0.45\textwidth]{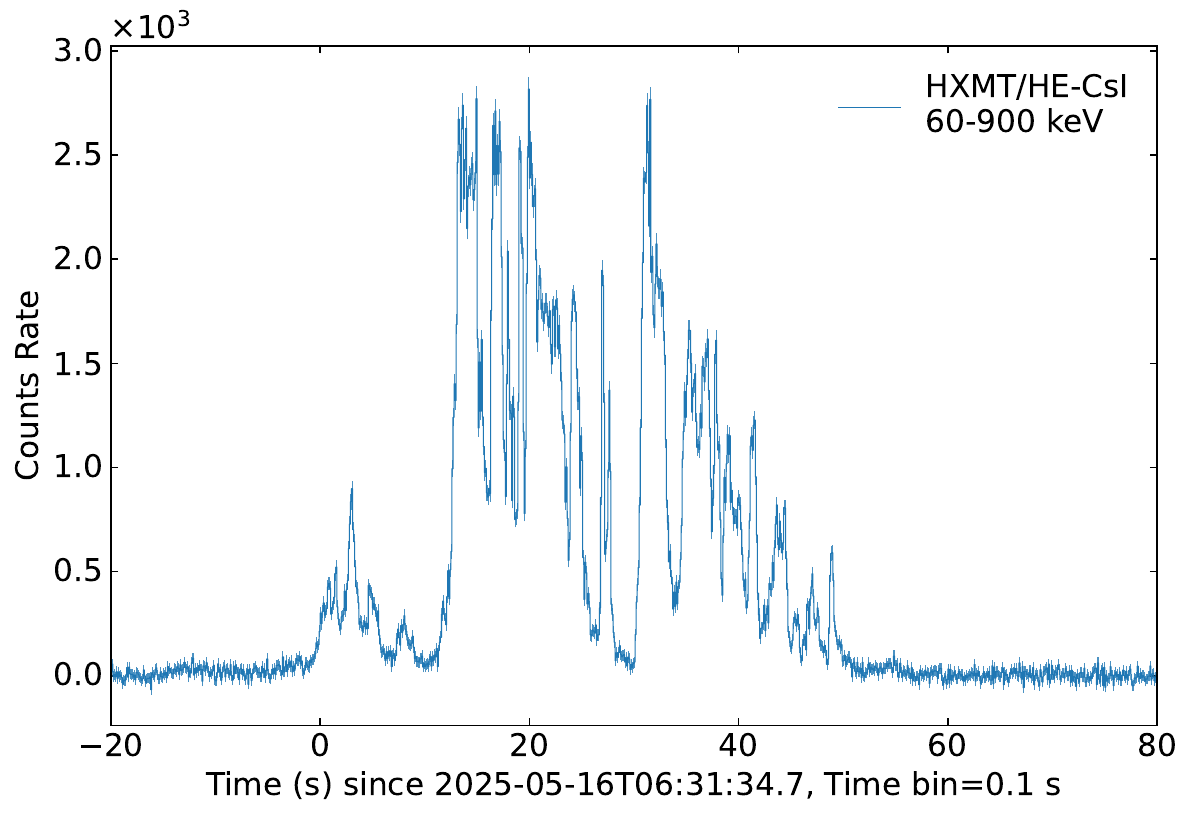}\put(-3, 65){\bf b}\end{overpic} \\
\begin{overpic}[width=0.45\textwidth]{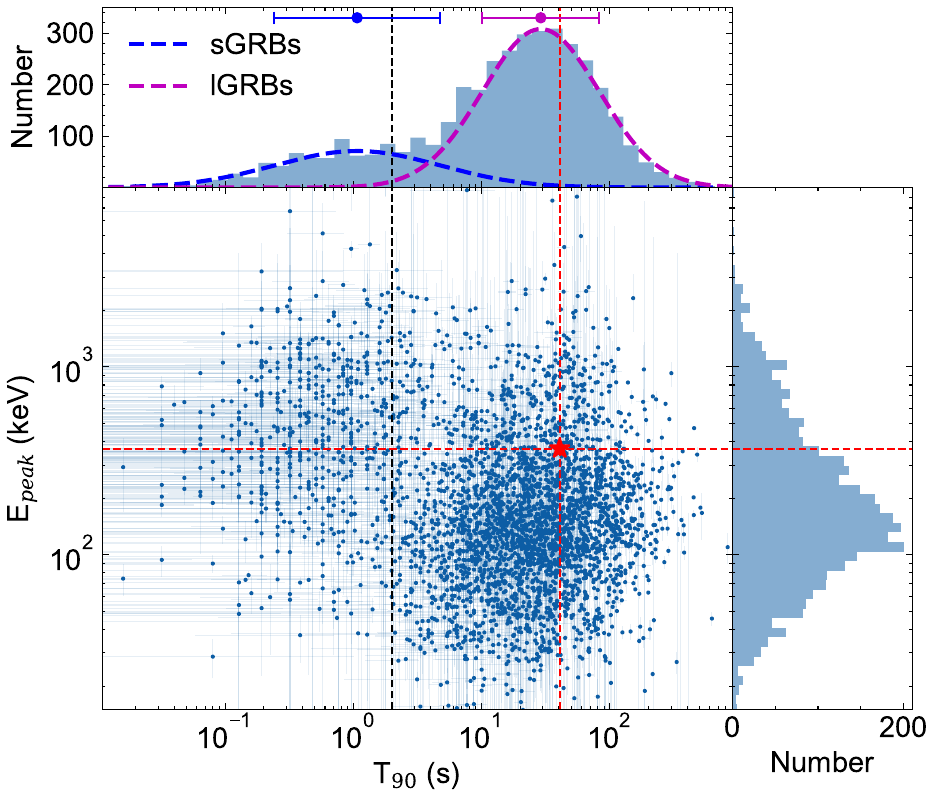}\put(-3, 80){\bf c}\end{overpic} &
        \begin{overpic}[width=0.45\textwidth]{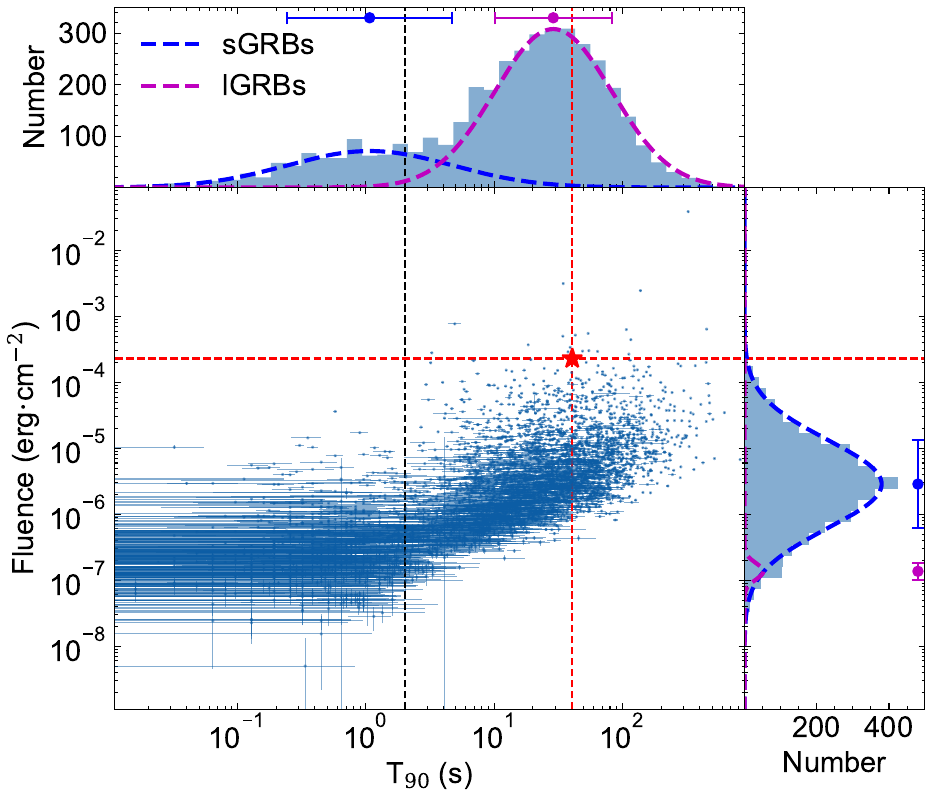}\put(-3, 80){\bf d}\end{overpic}  \\
\begin{overpic}[width=0.45\textwidth]{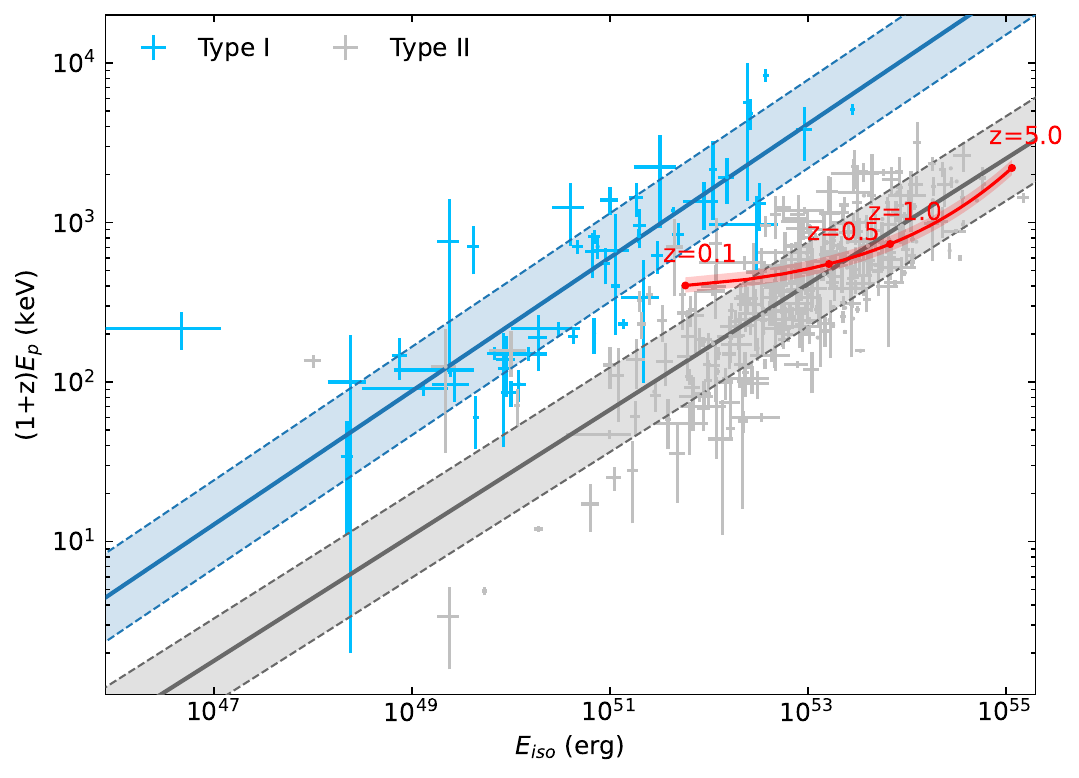}\put(-3, 65){\bf e}\end{overpic} &
        \begin{overpic}[width=0.45\textwidth]{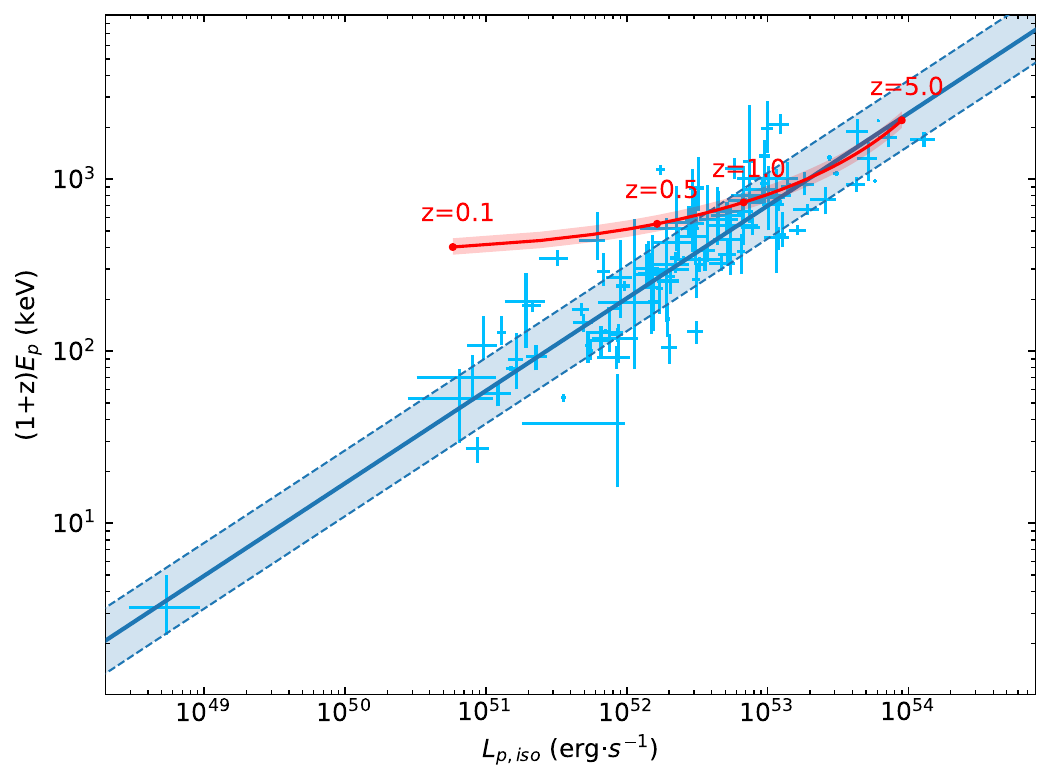}\put(-3, 65){\bf f}\end{overpic}     
\end{tabular}
\caption{\noindent\textbf{Part of \texttt{BREAKFAST} analysis results of bright burst GRB 250516B.} 
(a) and (b), the GECAM-B and HXMT lightcurve of GRB 250516B. 
(c) and (d),  the localization of GRB 250516B on the diagram of $T_{90}$-E$_{\rm peak}$ and $T_{90}$-Fluence diagram. 
(e) and (f), the localization of GRB 250516B on the Amati relation diagram and Yonetoku relation diagram.}
\label{fig:0516}
\end{figure*}

An event that reflects the important role of \textit{GHS} and \texttt{BREAKFAST} in TDAMM is GRB 250516B, which is first detected and reported by GECAM-B and processed by \texttt{BREAKFAST} \citep{250516B_GECAM_GCN}. 
In \textit{GHS}, another instrument, HXMT, also detected this bright GRB. 
Some of the \texttt{BREAKFAST} results are shown in Figure\,\ref{fig:0516}. 
The IPN format data is quickly produced by \texttt{BREAKFAST} and plays an important role in IPN localization of this burst \citep{250516B_IPN_GCN}. 

\begin{figure*}
\centering
\begin{overpic}[width=0.9\textwidth]{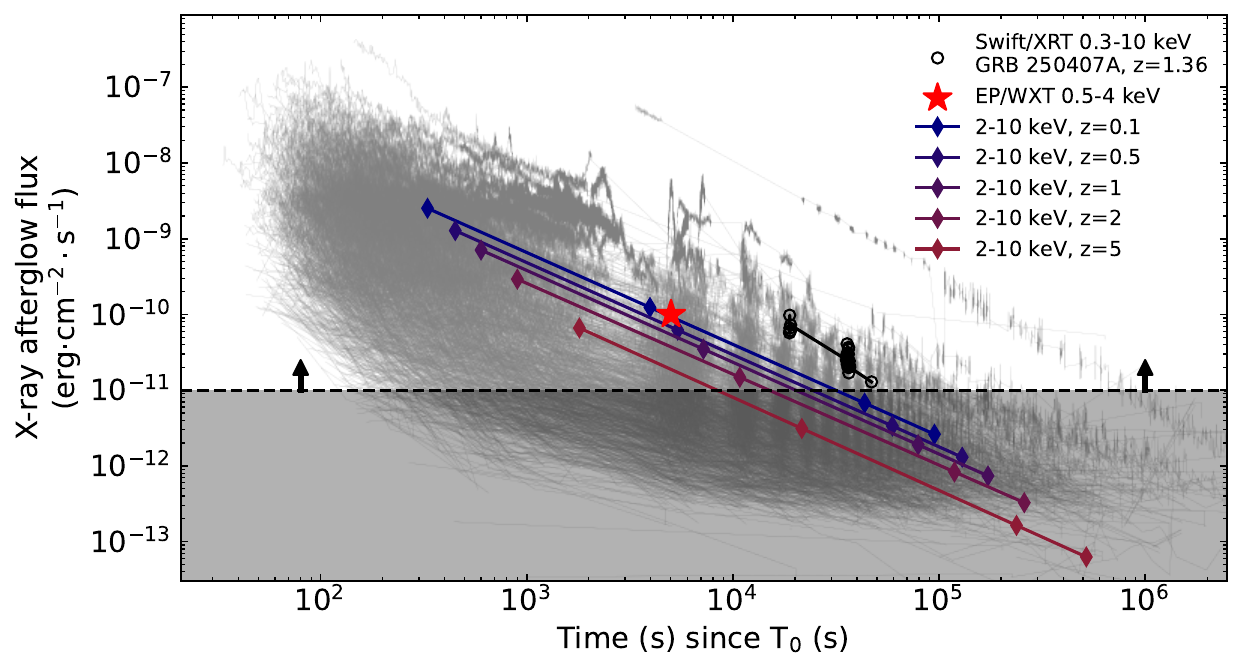}\put(-3, 50){\bf a}\end{overpic} \\
\begin{overpic}[width=0.9\textwidth]{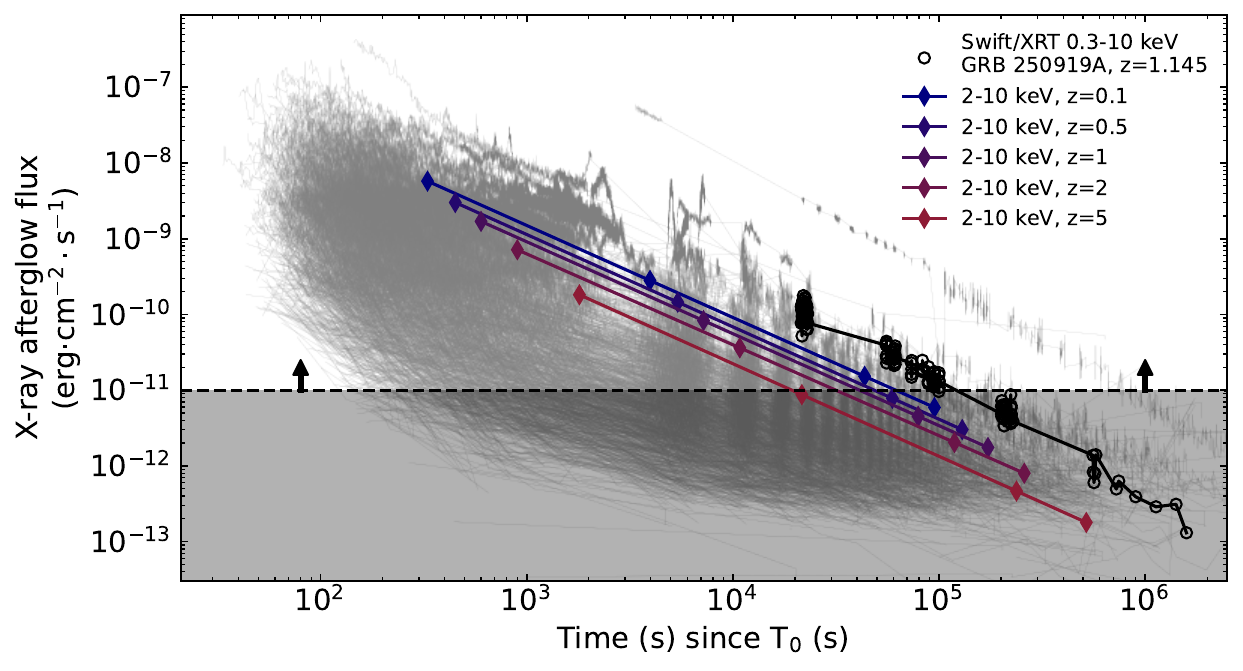}\put(-3, 50){\bf b}\end{overpic} 
\caption{\textbf{X-ray afterglow prediction result of two EP/WXT-afterglow-detected GRB 250407A and GRB 250919A.} 
The solid line in blues and reds are the flux lightcurve prediction of X-ray afterglow based on observation prompt emission with different redshift assumptions. 
Each line of the flux lightcurve prediction has four diamond-shaped points, which represent the time since of 300 seconds, 1 hour, 11 hours, and 1 day after the trigger in the rest frame. 
The black circles are the X-ray afterglow detected by \textit{Swift}/XRT. 
The background gray line is all the X-ray afterglow flux lightcurve in the observer frame detected by \textit{Swift}/XRT \citep{xrt_catalog_1,xrt_catalog_2}. 
The black dashed line represents the sensitivity of EP/WXT, which is about $10^{-11} \rm \, erg \cdot cm^{-2} \cdot s ^{-1}$ in 0.4-5\,keV. 
The red star in (a) is the unabsorbed 0.5-4 keV flux of GRB 250407A afterglow detected by EP/WXT \citep{250407A_EP_GCN}. 
The spectral parameters of prompt emission is the same with \cite{250407A_GRM_GCN} and \cite{250919A_GRM_GCN}.}
\label{fig:Fig_ag}
\end{figure*}

Two successful examples of afterglow predictions are the SVOM/GRM detected bursts GRB 250407A \citep{250407A_GRM_GCN} and GRB 250919A \citep{250919A_GRM_GCN}, which was also detected by EP/WXT \citep{250407A_EP_GCN,250919A_EP_GCN}. 
The reported flux of X-ray afterglow by EP/WXT is well consistent with the prediction by \texttt{BREAKFAST}, as depicted in Figure\,\ref{fig:Fig_ag}(a)
The X-ray afterglow of GRB 250919A was first identified as a new X-ray source EP250919a, as the transient was detected two hours after the trigger of GRB; but EP250919a is subsequently confirmed as the X-ray afterglow of GRB 250919A. Although information of the flux measured with EP/WXT is not publicly available yet, the X-ray afterglow detection of EP/WXT is well consistent with the predictions by \texttt{BREAKFAST} that the X-ray afterglow flux should be brighter than 10$^{-10}$ erg$\cdot$cm$^{-2}$$\cdot$s$^{-1}$ at two hours later than the trigger time, as shown in Figure\,\ref{fig:Fig_ag}(b).

\subsection{High-z GRBs}
GRB 250314A is a high-z GRB with a redshift of $z\sim$7.3 \citep{250314_SVOM}, the $E_{\rm iso}$ is first reported by GRM with the help of \texttt{BREAKFAST} \citep{250314_GRM_GCN}. 
Similarly, \texttt{BREAKFAST} is also applied in the GHS observations of other intermediate or high redshift GRBs, including GRB 250215A \citep{250215_GRM_GCN} and GRB 250725A \citep{250725_GRM_GCN}. 

Moreover, by combining predictions and observations of the X-ray afterglow, certain constraints can be placed on the redshift of GRBs. Specifically, if the analysis based on the prompt emission by \texttt{BREAKFAST} predicts that the X-ray afterglow can be detected by EP/WXT within a certain time window for redshifts below 5, but EP/WXT did not actually detect the X-ray counterpart within the corresponding time windows, this suggests that the redshift of the GRB is very likely greater than 5. 
Although the luminosity of X-ray afterglow is influenced by various factos (e.g. circumburst density, bulk Lorentz factor, et al.), this still provides a feasible approach to filtering the possible high-z GRB.

\subsection{Merger-origin GRBs}
The merger-origin GRBs are currently the research emphasis of TDAMM observations, as they are potential Gravitational Wave Electromagnetic counterparts (GWEMs).

As an interesting subclass of merger-origin GRBs, Type IL GRBs (such as GRB 211211A and GRB 230307A) feature a long-duration prompt emission \citep{Type_IL_1,Type_IL_2}. 
A three-episode burst pattern of Type IL GRB was proposed to identify this peculiar burst quickly. 
The burst pattern is also used for searching Type IL GRBs in \texttt{BREAKFAST}, and two events were identified: GRB 250521D \citep{0521D_GRM_GCN} and GRB 250819A \citep{0819A_GRM_GCN} and follow-up observations were conducted by SVOM/VT. 

Additionally, several short GRBs with extended emission were identified by \texttt{BREAKFAST}, including GRB 241117A \citep{241117A_GRM_GCN} and GRB 250912A \citep{250912A_GRM_GCN}.

\subsection{Magnetar-associated phenomena}
GRB 241107A is an extragalactic magnetar giant flare candidate, which was first detected and reported by SVOM/GRM \citep{241107A_GRM_GCN}. 
The IPN format data is quickly produced by \texttt{BREAKFAST} and utilized in IPN joint localization \citep{241107A_IPN_GCN}, which guided a series follow-up observations, including SVOM/VT and EP/FXT. 

Another possible extragalactic magnetar giant flare candidate is GRB 250308A, which was also first detected and reported by SVOM/GRM \citep{0308A_GRM_GCN_1}. 
Interestingly, the QPO analysis result from \texttt{BREAKFAST} suggests a possible QPO signal in the burst \citep{0308A_GRM_GCN_2}. 
The \texttt{BREAKFAST} \citep{0308A_IPN_GCN}, and a series of follow-up observations with EP/FXT, SVOM/VT, as well as the KM-40M telescope, were conducted. 

Additionally, the discovery of FRB/MXB 200428 \citep{HXMT_200428,CHIME_200428,START2_200428} reveals that the magnetar is one of the origins of FRB. More FRB-associated MXBs were detected with GECAM-B \citep{GECAM_221014_ATEL}. The joint observation of \textit{GHS} provides longer coverage time than a single instrument, therefore it is expected to obtain a more comprehensive catalog of MXBs. \texttt{BREAKFAST} has already provided quick analysis for many MXBs, including bursts from SGR 1806-20 \citep{1806_GCN_1,1806_GCN_2}, Swift J1818.0-1607 \citep{J1818_GCN_1} and SGR 1E 1841-045 / Kes 73 \citep{1841_GCN_1,1841_GCN_2}. 
Once the rate of MXBs has increased, the joint localization will be conducted to confirm their origins, and multi-wavelength follow-up telescopes, especially radio telescopes, will be promptly notified to monitor potential FRB-associated MXBs.

\subsection{Upper limit of other TDAMM transients} 
\texttt{BREAKFAST} also used to providing upper limit in $\gamma$-ray band for transients trigger at other wavelengths (e.g., EP240408a \citep{0408_upper_GCN}, EP240425a, EP241217a, et al.) and other messenger, including high energy neutrino (e.g., GRB 250309B/IceCube-250309A \citep{0309B_GECAM_GCN}), and GW (e.g., S250206dm \citep{S250206dm_LVK_GCN, S250206dm_GECAM_GCN}).

\section{Discussion and Conclusion} \label{sec:summary}
As an important part of the TDAMM joint observation, the \textit{GHS} constellation provided a new multi-instrument joint BA duty paradigm for other wide FoV $\gamma$-ray monitors. 
To reduce the workload of BAs and improve the efficiency of joint observations, we developed a new pipeline framework for \textit{GHS} joint BA duty, by extending and integrating available pipeline. 
We also modified the multi-instrument joint data analysis pipeline \texttt{ETJASMIN} to specifically adapt to the needs of BAs' works.

\texttt{BREAKFAST} aims to providing quick, preliminary but standard data reduction results for BAs. 
As a framework with clear objectives and requirements, \texttt{BREAKFAST} will not provide in-depth physical analysis, such as fitting spectra by physical models and multi-instrument joint spectrum fitting, nor does it pursue excessive flexibility, meaning that not all parameters are made visible or adjustable to BAs; otherwise, the complexity of learning and operation would increase significantly, as there will always be special cases that cannot be solved through standard procedures. 

More importantly, \texttt{BREAKFAST} is also the first framework designed and applied for observation and analysis for joint duty across multiple instruments and constellations. 
Therefore, this framework, which unifies the processing of triggers between different instruments through top-level design, also provides experience for collaboration among other $\gamma$-ray monitor constellations in the future.

\texttt{BREAKFAST} has not yet reached a perfect state and requires an amount of upgrading and improvement work. 
For example, although the X-ray afterglow prediction is achieved in \texttt{BREAKFAST}, and there are also many optical and radio follow-up telescopes that maintain close contacts with \textit{GHS}, especially wide FoV optical telescope array GWAC, however, the prediction of the brightness of optical afterglow and radio afterglow is not implemented in \texttt{BREAKFAST} yet. Exploration of the correlation between optical/radio afterglow luminosity and prompt $\gamma$-ray emission properties, as well as building such a module in \texttt{BREAKFAST}, will be one of the main subjects for the upgrading work in the future. 

In addition, at the current stage, the selection of some experience-based parameters has led to a limited success rate for the automatic analysis process (such as the background time interval), necessitating a semi-automatic process by resetting some parameters. 
An iterative upgrade of adaptive parameter settings through the introduction of some machine learning techniques is also underway, which will be implemented and reported in the forthcoming work. 

Nevertheless, at the time of this writing, a considerable number of joint observation campaigns associated with \texttt{BREAKFAST} have shown that this framework, as well as \textit{GHS} joint BA duty, has performed effectively. 
The inclusion of more instruments into the joint BA duty and \texttt{BREAKFAST} framework has been put on the agenda.

Currently, \texttt{BREAKFAST} has invoked in the data processing of over one hundred GCN circulars. 
As a service for \textit{GHS} BAs on duty, hoping that the \texttt{BREAKFAST} with eating Jasmine (\texttt{ETJASMIN}) will help to start a wonderful day for BAs. 
In the future, \texttt{BREAKFAST} also plans to gradually open up to the TDAMM community through AstroBurstHub (within the scope permitted by data policies), becoming an official repository for prompt emission observation and analysis of \textit{GHS} constellation.

\begin{acknowledgments}
We appreciate Yu Wang, Li-Ping Xin and Hong-Mei Zhang for the helpful discussion and suggestions that improved this work. 
This work is supported by 
the National Natural Science Foundation of China (Grant No. 12494572, 
12273042
),
the National Key R\&D Program of China (2021YFA0718500),
the Strategic Priority Research Program, the Chinese Academy of Sciences (Grant No. 
XDA30050000,
XDB0550300
).
This work made use of data from the \textit{Insight}-HXMT mission, funded by the CNSA and CAS.
The GECAM (Huairou-1) mission is supported by the Strategic Priority Research Program on Space Science (Grant No. XDA15360000) of Chinese Academy of Sciences. 
The Space-based multi-band Variable Objects Monitor (SVOM) is a joint Chinese-French mission led by the Chinese National Space Administration (CNSA), the French Space Agency (CNES), and the Chinese Academy of Sciences (CAS). We gratefully acknowledge the unwavering support of NSSC, IAMCAS, XIOPM, NAOC, IHEP, CNES, CEA, and CNRS.
We are grateful to the development and operation teams of \textit{Insight}-HXMT, GECAM and SVOM. 
We appreciate the public data and software of \textit{Fermi}/GBM. 
This work made use of data supplied by the UK Swift Science Data Centre at the University of Leicester. 
We show our profound gratitude to all the Burst Advocates who participated in the joint duty of GECAM-HXMT-SVOM, as well as all the telescope teams conducting follow-up observations and joint observations.
\end{acknowledgments}

% \begin{contribution}
% % All authors contributed equally to the Terra Mater collaboration.
% \end{contribution}

\clearpage

\bibliography{main.bbl}{}
\bibliographystyle{aasjournalv7}

\end{document}